\documentclass[twocolumn,prl,aps,showpacs,footinbib,superscriptaddress,longbibliography]{revtex4-1}
\usepackage{array}
\usepackage{graphicx}
\usepackage{dcolumn}
\usepackage{color}
\usepackage{bm}
\usepackage{natbib}
\usepackage{hyperref}
\usepackage{amsmath,amssymb,amsfonts,bbold}
\usepackage[normalem]{ulem}

\usepackage{lineno}
\usepackage{miller}
\usepackage{tikz}
\usepackage{physics}

\newcommand{\beq}{\begin{equation}}
\newcommand{\eeq}{\end{equation}}
\newcommand{\beqa}{\begin{eqnarray}}
\newcommand{\eeqa}{\end{eqnarray}}

\begin{document}
\preprint{}
\title{Observation of zero-energy modes in Gd atomic chains on superconducting Nb(110)}

\author{Yu Wang}
\email[]{yu.wang@physik.uni-wuerzburg.de}
\author{Felix Friedrich}
\affiliation{Physikalisches Institut, Experimentelle Physik II,
Universit\"at W\"urzburg, Am Hubland, 97074 W\"urzburg, Germany}
\author{Matthias Bode}
\affiliation{Physikalisches Institut, Experimentelle Physik II,
Universit\"at W\"urzburg, Am Hubland, 97074 W\"urzburg, Germany}
\affiliation{Wilhelm Conrad R\"ontgen-Center for Complex Material Systems (RCCM), Universit\"at W\"urzburg, Am Hubland, 97074 W\"urzburg, Germany}
\author{Artem Odobesko}
\affiliation{Physikalisches Institut, Experimentelle Physik II,
Universit\"at W\"urzburg, Am Hubland, 97074 W\"urzburg, Germany}
\date{\today}

\begin{abstract}
In this experimental study, we use scanning tunneling microscopy and spectroscopy to investigate Yu-Shiba-Rusinov states induced by $4f$-shell rare-earth Gd adatoms on a superconducting Nb(110) surface. We engineer Gd atom chains along the substrate’s \hkl[1-10] and \hkl[001] directions, revealing distinct behaviors in differently oriented chains. \hkl[1-10]-oriented Gd chains exhibit spectroscopic features at their ends, identifying them as trivial edge states, while \hkl[001]-oriented Gd chains display zero-energy edge states, suggesting non-trivial nature. Notably, Gd chains with four atoms--independent of their particular orientation--exhibit a uniform zero-energy mode along the entire chain. These findings call for further research and a theoretical framework to describe rare-earth-based structures on superconductors.

\end{abstract}


\maketitle

\section{Introduction}

The presence of a magnetic impurity in a superconductor leads to the emergence of bound states referred to as Yu-Shiba-Rusinov (YSR) states~\cite{Yu1965, Shiba1968, Rusinov1969}. These states, extensively reviewed in Ref.~\cite{Balatsky2006}, can be probed using scanning tunneling spectroscopy (STS) and manifest themselves as narrow resonances in tunneling spectra~\cite{Heinrich2018}. The proposal to generate topologically protected edge states by linking YSR states within a one-dimensional(1D) chain of magnetic adsorbates \cite{Braunecker2013, NadjPerge2013, Klinovaja2013a, Pientka2013} has sparked considerable interest in this field. Numerous research studies have been conducted on $3d$ transition metals placed on various superconducting substrates, ranging from single impurities~\cite{ruby2016,choi2018, Odobesko2020, Schneider2019, Menard2015, Kuester2021, Cornils2017} to dimers~\cite{Ji2008, Beck2021, beck2023,friedrich2021,ruby2018} and long 1D chains~\cite{pawlak2016,nadj2014,ruby2017,kim2018,liebhaber2022,mier2021,schneider2021,schneider2022precursors, Kamlapure2018}. 

In contrast, experimental results for $f$-shell rare-earth metals (REMs), such as Gd, Tb, Dy, or Ho, are scarce and essentially limited to pioneering work on single Gd atoms on Nb(110) surface~\cite{Yazdani1997} and Gd atoms on Bi(110) films grown on Nb(110)~\cite{Ding2021}. 
However, REMs on superconductors are highly intriguing owing to their magnetic properties, which fundamentally differ from those of the $3d$ transition metal elements: firstly, the magnetic moment of REMs primarily originates from the highly localized $4f$ orbital and is screened by outer electronic shells. Consequently, there is no direct interaction with $4f$ magnetic moment, but instead, the interaction is mediated indirectly through the outer $5d$ and $6s$ orbitals ~\cite{Wienholdt2013, Pivetta2020}. Secondly, the presence of higher spins, strong single-ion anisotropy, indirect Ruderman-Kittel-Kasuya-Yosida coupling as well as Dzyaloshinsky-Moriya interactions are more characteristics of REMs, thereby atomic chains constructed from these elements potentially enable a novel route towards helical-spin chains on superconductors exhibiting Majorana bound states, as proposed in  Refs.~\cite{Braunecker2013, Klinovaja2013, martin2012, Vazifeh2013, Schmid2022}. 

To initiate this research path, in this paper we present a hybrid system configuration comprised of Gd adatoms on a superconducting Nb\mbox{(110)} surface. We observe that individual Gd adatoms on Nb\mbox{(110)} do not exhibit any YSR in-gap states. However, we find that there are two distinct nearest neighbor arrangements of two Gd atoms, along the \hkl[1-10] and \hkl[001] direction of the substrate, which display hybridized YSR in-gap states. We are able to build close-packed chains from Gd atoms along these specific directions and demonstrate that \hkl[1-10]-oriented chains exhibit a spectroscopic feature at the chain ends akin to trivial edge states, whereas \hkl[001]-oriented chains built from four or more Gd atoms exhibit zero-energy end states.

\section{Experimental methods}

All measurements are conducted using a custom-made low-temperature scanning tunneling microscope (STM) which is operated at tip and sample base temperature of 1.4\,K. The Nb\mbox{(110)} crystal is cleaned through sputtering with Ar ions and a series of high-temperature annealing cycles~\cite{Odobesko2019}. Gd atoms are deposited \textit{in-situ} onto the cold Nb\mbox{(110)} substrate at a temperature of 4.2\,K. Spectroscopic measurements are conducted with lock-in technique with a modulation voltage of 0.1\,mV at a modulation frequency of 890\,Hz. In order to enhance the energy resolution of the spectroscopic measurements, tungsten tips are functionalized 
with a Nb cluster on the tip apex (see Ref.~\cite{Odobesko2020} for details of the experimental procedure). This results to the shift of all spectral features by the value of the superconducting(SC) gap of the tip $\Delta_\mathrm{tip}$, indicated in all figures by dashed lines.

\section{Results}

\begin{figure}
\centering
	\includegraphics[width=\linewidth]{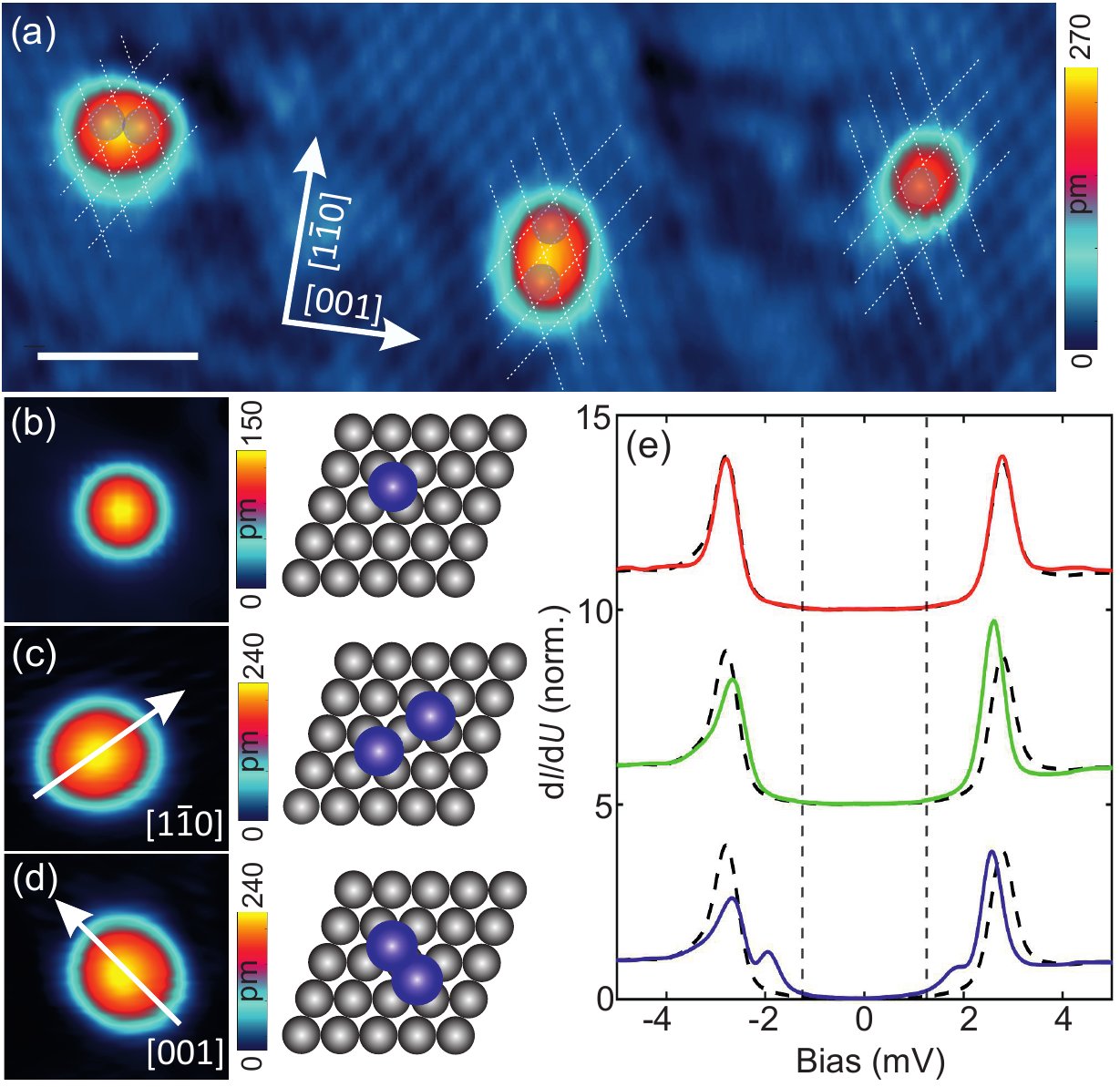}
	\caption{Topographic images of Gd adatoms on a clean Nb(110) surface in constant-current mode. 
		\mbox{(a)} atomic resolution of the \mbox{(110)} surface with self-assembled Gd structures recorded with a CO-functionalized tip.
		\mbox{(b)} single Gd atom  and \mbox{(c,d)} assembled Gd dimers recorded with SC-functionalized tip along with corresponding ball models. The white scale bar is 2 nm and applies to all topographic images in this figure.
		\mbox{(e)} Tunneling spectra of single Gd adatom and the two Gd dimers. The black hatched lines represent the spectra on clean Nb(110).  Spectra are vertically shifted for clarity. The gap of the superconducting tip amounts to \mbox{$\Delta_\mathrm{tip} = 1.25\,\mathrm{meV}$} and is indicated by vertical gray dotted lines. The orientations of the dimers are indicated in each panel. 
		Stabilization parameters: \mbox{(a) $U = -7\,\mathrm{mV}$,} \mbox{$I = 1.5\,\mathrm{nA}$;} 
		\mbox{(b)-(d) $U = -20\,\mathrm{mV}$,} \mbox{$I = 400\,\mathrm{pA}$;} 
		\mbox{(e) $U = -7\,\mathrm{mV}$,} \mbox{$I = 400\,\mathrm{pA}$.}  }
	\label{dimers}
\end{figure}

Figure 1(a) shows a constant-current STM image of the Nb(110) surface with priorly deposited Gd adatoms, captured using a CO-functionalized tip to increase spatial resolution~\cite{friedrich2021}. The atomically resolved body-centered cubic (110) structure of the Nb surface 
with a lattice constant of $a = 330$\,pm can be clearly seen. Although, the presence of CO significantly increases the spatial resolution, the extension of Gd atoms shown as a bright protrusions in the topographic map considerably exceeds the size of the niobium unit cell. This large apparent size is presumably caused by the relatively large radius of the $6s$-orbital of Gd, that makes it challenging to unambiguously determine the position of adsorbed atoms within the Nb unit cell. Nevertheless, considering the surrounding crystalline structure as shown in the Fig.~\ref{dimers}(a), we can infer the probable adsorption position of the Gd atoms to be the fourfold coordinated hollow site of the Nb(110) lattice.

In the left part of Fig.~\ref{dimers}(a), two dimers self-assembled during the deposition process along the \hkl[1-10] and \hkl[001] directions of the substrate can be recognized. The enhanced resolution provided by the CO-functionalized tip reveals a distinctly elongated shape along the dimer axis. By using atomic manipulation and displacing Gd atoms exclusively along the given atomic assembly direction, we can intentionally create dimers along the \hkl[1-10] and \hkl[001] directions of the substrate, with exactly the same spectral characteristics as self-assembled ones. In contrast to the topographic images obtained with a CO-tip, the elongated shape of the dimers along the assembly direction (indicated with arrows in Fig.~\ref{dimers}(c-d) ) obtained by a superconducting tip is weakly distinguishable.


A corresponding ball model of the single Gd atom and the two assembled Gd dimers in the \hkl[1-10] and \hkl[001] direction are presented in Fig.~\ref{dimers}(b-e) along with characteristic tunneling spectra.  Considering that the single Gd atom is adsorbed in a fourfold hollow adsorption site, we anticipate that, due to the Gd--Gd interatomic interactions, the interatomic distance in dimers will likely experience a significant inward or outward displacement~\cite{bode1996nanostructural,friedrich2021}. Comparison of the nearest neighbor distance of bulk Gd, $a_{\rm Gd} = 363.6\,\mathrm{pm}$ with the Nb\mbox{(110)} lattice parameters, $a_{\rm Nb\hkl[1-10]} = 376.7\,\mathrm{pm}$ and $a_{\rm Nb\hkl[001]} = 330\,\mathrm{pm}$, leads one to expect that the \hkl[1-10]-oriented Gd dimer is under compressive strain, whereas the \hkl[001]-oriented Gd dimer is expansively strained, likely resulting in an inward and outward relaxation, respectively.

Tunneling spectra atop of a single Gd atom do not show any in-gap states and are indistinguishable from the spectra taken on clean Nb(110), see red and black hatched lines in Fig.~\ref{dimers}(e), respectively. Further, we compare representative spectra of the \hkl[1-10]-oriented dimer, which has the larger interatomic distance [cf.\ Fig.~\ref{dimers}\mbox{(e)} green line], with the spectrum of the bare Nb(110) surface. A small deviation of the $\dd I/\dd U$ signal at the position of the coherence peaks, with a more (less) pronounced resonance at positive (negative) energy for the dimer can be recognized. The presence of only one pair of YSR states close to the superconducting gap edges indicates a weak hybridization of YSR states within the dimer. Hybridization of individual YSR states should result in energy splitting of the resonances~\cite{friedrich2021, choi2018,ruby2018}, which in case of the \hkl[1-10]-oriented dimer is either absent or is beyond our energy resolution.

\begin{figure*}
    \centering
	\includegraphics[width=\linewidth]{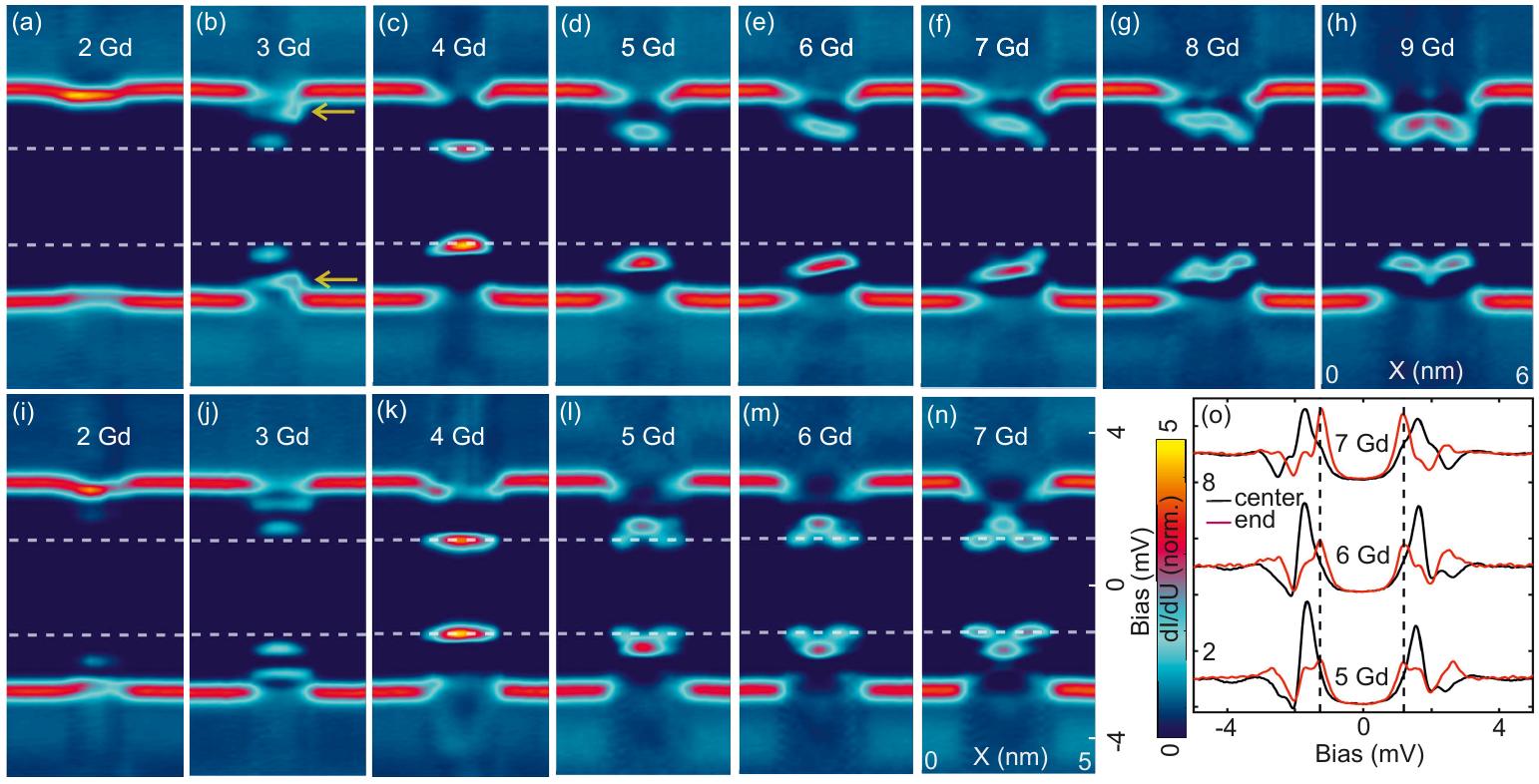}
	\caption{\mbox{(a-h)} The $\dd I/\dd U$ signal measured along the \hkl[1-10]-oriented chains consisting of 2 -- 9 Gd atoms. 
		\mbox{(i-n)} The $\dd I/\dd U$ signal measured along the \hkl[001]-oriented chains consisting of 2 -- 7 Gd atoms. The scale of the line spectra is marked in \mbox{(n)}.  
		\mbox{(o)} Single point tunneling spectra of \hkl[001]-oriented chains consisting of 5 -- 7 Gd measured at the end(red) and at the center(black) of the chain. 
		\mbox{$\Delta_\mathrm{tip} = 1.25\,\mathrm{meV}$}. 
		Stabilization parameters: \mbox{ $U = -7\,\mathrm{mV}$,} \mbox{$I = 400\,\mathrm{pA}$.}   
	\label{fig:linegrids}}
\end{figure*}

In contrast, spectra of the \hkl[001]-oriented dimer, which has the smaller interatomic distance 
[cf. Fig.~\ref{dimers}\mbox{(e)} blue line] shows an additional pair of in-gap peaks. Similar to the \hkl[1-10]-oriented dimer, the more intense pair of resonances appears near the coherence peak, whereas the less intense pair of resonances appears at a bias voltage U=$ \pm (1.9 \pm 0.1)$ mV in the convoluted tip and sample local density of states (LDOS). In the next step, we used the atomic assembly method outlined above to construct Gd chains along these two specific directions.

In Fig.~\ref{fig:linegrids}(a-h), we present a series of spectroscopic measurements for densely-packed chains consisting of 2 to 9 Gd atoms along the \hkl[1-10] direction. The spectra are measured along a line following the chain axis and are presented in the form of a heat-map. While for two atoms, the YSR states are situated near the edges of the superconducting gap, see Fig.~\ref{fig:linegrids}(a), the addition of one more atom results in the emergence of a pair of YSR states far within the superconducting gap close to the Fermi level, see Fig.~\ref{fig:linegrids}(b). 

Notably, the direction from which the atom was added to the chain is clearly discernible in the $\dd I/\dd U$ line spectra. For example, in the case of 3-atom chain there is a distortion in the heat-map spectrum on the right side of the chain, near the coherence peaks, precisely at the end where the atom was attached, see arrow in Fig.~\ref{fig:linegrids}(b). A similar behavior is also observed for 6-, 7-, and 8-atom \hkl[1-10]-oriented chains, where a distortion is always noticeable in the YSR states near the Fermi level, see Fig.~\ref{fig:linegrids}(e)-(g). This distortion is likely a consequence of an inhomogeneous, gradient-like strain that arises during the assembly of densely-packed chains and which cannot fully relax without external stimuli. In comparison to the relaxed bulk of the chain, this residual strain may cause a weak distortion of the atom-atom or atom-substrate coupling.

However, starting from 9-atom chains, this asymmetry disappears, indicating that for longer chains the strain is evenly distributed along the chain, and the spectra are symmetric with respect to the chain center, see Fig.~\ref{fig:linegrids}(h). A closer examination of the position of the YSR states for 2- to 9-atom chains unequivocally highlights the unique properties of the 4-atom chain from all other configurations. While in all other cases the YSR states are located within the gap and only approach the Fermi level towards the chain end, the YSR states of the 4-atom chain are energetically located at the Fermi level for the entire chain, i.e., not only at the two ends of the 4-atom chain but also in their ``bulk''. Moreover, there are no indications of any strain gradient, as previously described for the other shorter chains. This remarkable behavior of the 4-atom chain will also be discussed for Gd chains aligned along the [001] direction below.

Gd chains assembled along the \hkl[001] direction exhibit striking differences as compared to \hkl[1-10]-oriented chains, as presented in the bottom row of Fig.~\ref{fig:linegrids}(i)-(n). 
For a 3-atom chain, the spectrum represented in a heat-map shows two pairs of YSR states, similar to the dimer, cf.\ Fig.~\ref{fig:linegrids}(i)-(j), but more pronounced and shifted closer to the Fermi level. It is particularly noteworthy that for the 4-atoms chain all YSR states suddenly reappear at the Fermi level and distribute uniformly along the entire chain (Fig.~\ref{fig:linegrids}(k)), as in the case of the \hkl[1-10]-oriented chain. Subsequent addition of another Gd atom leads to the formation of distinct edge states which are energetically located at the Fermi level and bulk states that are shifted away from the Fermi level by approximately 0.2 meV. Please note, that already for the 5-atom long \hkl[001]-oriented Gd chain the edge states are energetically separated from the bulk ones Fig.~\ref{fig:linegrids}(l), in contrast to the \hkl[1-10]-oriented chain, where a smooth transition from bulk to edge states is observed, see Fig.~\ref{fig:linegrids}(h). Further increasing the chain length does not lead to significant changes in the spectral features,  
as evidenced by the data of 6- and 7-atom chains, see Fig.~\ref{fig:linegrids}(m) and (n), respectively. Also for these longer chains the edge states at the Fermi level are energetically separated from the higher energy of bulk states.

It would be reasonable to expect the observation of quasi-particle interference (QPI) along the chain, signifying the existence of a one-dimensional bulk band formed through the hybridization of individual YSR states. This phenomenon has been documented in 1D chains formed from Mn or Cr atoms on Nb(110), for example~\cite{schneider2021, kuster2022non}. However, the short length of the chain may prevent the observation of any QPI patterns. Attempts to increase the length of \hkl[001]-oriented chains beyond eight atoms led to the destruction of the linear order of the chain, probably due to the accumulation of compressive strain. Our results indicate that for a chain length of more than eight atoms, the chains start to develop a zig-zag structure or even break into two smaller parts, which is accompanied by the disruption of the characteristic spectral features.

\section{Discussion}
We first discuss the absence of YSR states for an individual Gd atom on the pristine Nb(110) surface. Surprisingly, this is in contrast to the previous report~\cite{Yazdani1997}. The discrepancy may be attributed to the use of different substrates. In the case of ~\cite{Yazdani1997}, it is an oxygen-reconstructed Nb surface ~\cite{Odobesko2019}. As demonstrated in Ref. \cite{Odobesko2020}, the presence of oxygen can significantly alter the degree of hybridization of magnetic impurity with the superconducting substrate, which can lead to the emergence of YSR states. However, our measurements on single Gd adatoms on an oxygen-reconstructed Nb surface(Fig. S4 of the  Supplementary Material) have revealed controversial data with the findings in Ref.~\cite{Yazdani1997}. 
The absence of YSR states for an individual Gd atom on Nb surfaces suggests that, despite the substantial magnetic moment of the half-filled Gd $4f$-shell (with spin of 7/2), it is effectively screened by the outer $5d$ and $6s$ electrons, apparently leading to an indirect interaction involving the outer orbitals.

Notable differences are also observed in the spatial maps of LDOS for the \hkl[001]-oriented Gd dimer, as shown in Figure. S5 of the Supplementary Material. Two pairs of hybridized YSR states do not display any even or odd symmetry for the combination of YSR wavefunctions, as expected for a magnetic dimer and as repeatedly observed for various $3d$ transition metals~\cite{friedrich2021,beck2023,choi2017}. It is also challenging to identify any indications of the atomic orbital shape responsible for the potential magnetic scattering channel ~\cite{ruby2016, kuster2022non}. The absence of these characteristics further underscores the complexity of the indirect interaction with the substrate through a combination of $5d6s$ orbitals, as also indicated in \cite{Pivetta2020}.

\begin{figure}
\centering
	\includegraphics[width=\linewidth]{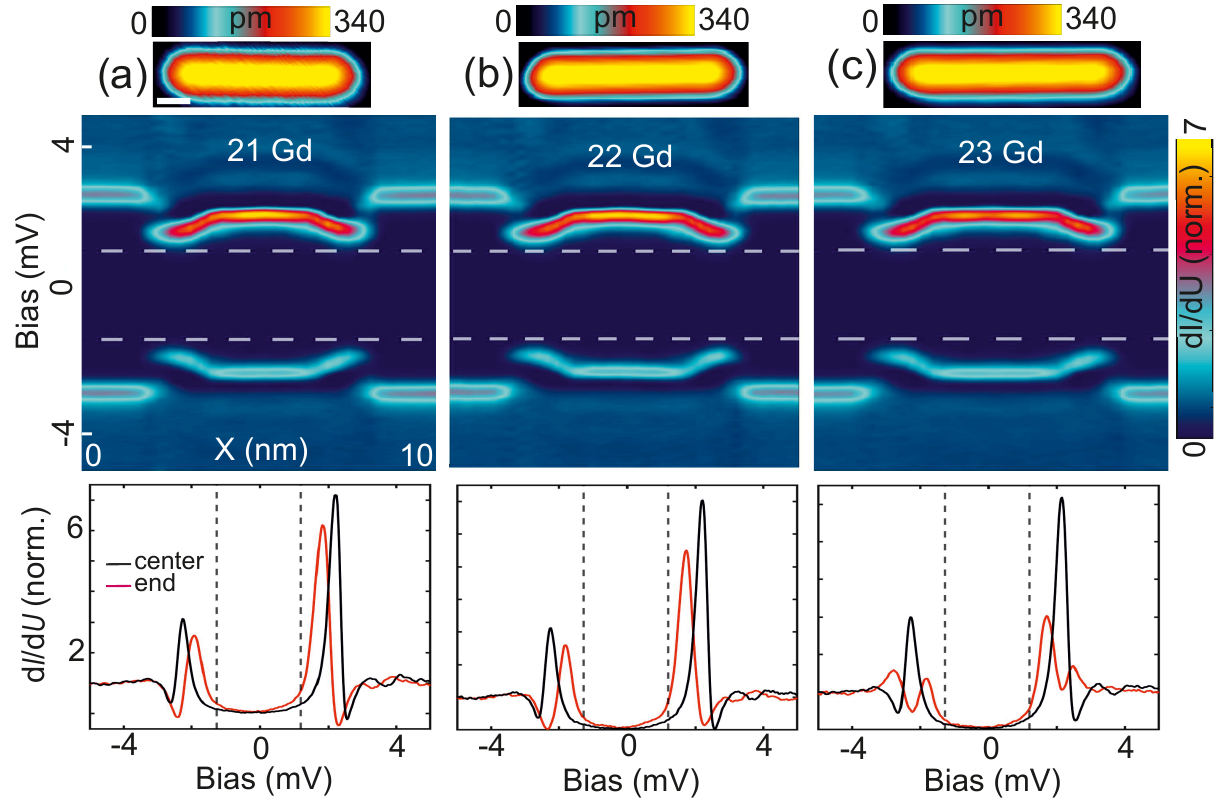}
	\caption{\mbox{(a-c)} The top row shows topographic images of the \hkl[1-10]-oriented chains consisting of 21 -- 23 Gd atoms. The white scale bar is 1 nm. The middle row presents respective $\dd I/\dd U$ signals measured along chains. The tunneling spectra in the bottom row are measured at the end (red) and center (black) of the chains.
		\mbox{$\Delta_\mathrm{tip} = 1.25\,\mathrm{meV}$}. 
		Stabilization parameters: \mbox{(a-g) $U = -7\,\mathrm{mV}$,} \mbox{$I = 400\,\mathrm{pA}$. }
	\label{fig:longlinegrids}}
\end{figure}

Furthermore, it is important to emphasize the influence of strain within the chain, which causes a displacement of the Gd atoms from their 4-fold hollow site equilibrium positions. As discussed in Ref.~\cite{Odobesko2020}, the displacement of atoms with respect to the Nb(110) unit cell strongly affects the hybridization of orbitals of the magnetic atom and the superconducting host, consequently leading to variations in the energy of the YSR states. Based on these considerations, we anticipate that chains oriented along the \hkl[1-10] direction 
exhibit trivial edge states, resulting from strain along the chain which is partially released towards the chain ends. This is also corroborated by the findings presented in Fig.~\ref{fig:longlinegrids}, where---for longer Gd chains assembled along the \hkl[1-10] direction---the absence of a distinct boundary between the edge states and the bulk states is evident. The bulk YSR states of \hkl[1-10]-oriented chains gradually shift closer to the Fermi level for edge atoms but do not reach it. Secondly, the energetically narrow bulk YSR band and the absence of any observed QPI for the bulk YSR band imply a very weak hopping parameter within the \hkl[1-10]-oriented chain.

Conversely, chains along \hkl[001] direction imply the presence of potentially non-trivial topological edge states. First, these edge states are located at the Fermi level and are clearly energetically separated from the bulk states. Second, the position of these edge states is independent of the number of atoms in the chain, starting from 5 atoms. Despite this chain being even more distorted than the \hkl[1-10] chain, topological protection apparently forces the edge state to remain at zero energy symmetrically at both ends. We deliberately refrain from using the term ``Majorana states`` in this context, as it seems unlikely that potential Majorana states in such short chains would evade hybridization, leading to energy splitting~\cite{schneider2022precursors}. However, it should be noted, that the potential realization of helical spin order within such 1D chain may result in a larger effective topological superconducting gap compared to ferromagnetic or anti-ferromagnetic chains~\cite{schneider2022testing, beck2023}. Consequently, Majorana states may be more localized. Regrettably, our attempts to determine the spin order along \hkl[001]-oriented Gd chains did not yield definitive results, presumably due to the very short length of such chains.

The last point to discuss is the remarkable behavior of chains consisting of 4 Gd atoms. Regardless of the chain's orientation they exhibit the same intriguing spectral feature, i.e., a uniform YSR state at the Fermi level. Moreover, neither the presence of a strain gradient nor various hopping terms for differently oriented chains influence this spectral characteristic. The ``magical number'' of 4 atoms remains an enigma and requires further analysis and theoretical understanding of this intriguing behavior of $4f$-shell REM.

\section{Conclusion}
In our work, we study Gd atoms on superconducting Nb(110). Despite the absence of any YSR states for an individual Gd atom, Gd dimers along two \hkl[1-10] and \hkl[001] surface directions demonstrate the emergence of YSR states within the superconducting gap. For longer chains assembled along the \hkl[1-10] direction we find trivial edge states at the chain ends and for chains  assembled along the \hkl[001] direction potentially non-trivial topological edge states situated at the Fermi level. In the \hkl[001] direction we were unable to construct longer chains consisting of more than 8 atoms, as the accumulated strain within the chain cell disrupts the linear order and breaks the chain apart. 

\begin{acknowledgments}
The work was supported by the DFG through SFB 1180 (project C02). We acknowledge financial support by the Deutsche Forschungsgemeinschaft (DFG, German Research Foundation) under Germany's Excellence Strategy through W{\"u}rzburg-Dresden Cluster of Excellence on Complexity and Topology in Quantum Matter -- ct.qmat (EXC 2147, project-id 390858490).
\end{acknowledgments}

\bibliography{bibliography.bib}
\end{document}